\documentclass[12pt]{article}

\usepackage{amsmath}
\usepackage{amssymb}

\usepackage{graphicx}

\usepackage{cite}

\usepackage{color,framed}

\usepackage[labelfont=bf,labelsep=period,justification=raggedright]{caption}

\makeatletter
\renewcommand{\@biblabel}[1]{\quad#1.}
\makeatother

\date{}

\begin{document}

\begin{flushleft}
{\Large \textbf{Constructing backbone network by using tinker algorithm} }
\\
Zhiwei He$^{1,\ast}$, Meng Zhan$^{2}$, Jianxiong Wang$^{3}$, Chenggui Yao$^{1}$
\\
{$^1$} Department of Mathematics, Shaoxing University, Shaoxing, China
\\
{$^2$} State Key Laboratory of Advanced Electromagnetic Engineering and Technology, School of Electrical and Electronic Engineering,
Huazhong University of Science and Technology, Wuhan, China
\\
{$^3$} College of Science, Hubei University of Technology, Wuhan, China
\\

$^\ast$ Corresponding author. E-mail:hezhiwei10@mails.ucas.ac.cn
\end{flushleft}

\section*{Abstract}
Revealing how a biological network is organized to realize its function is one of the main topics in systems biology. The functional backbone network, defined as the primary structure of the biological network, is of great importance in maintaining the main function of the biological network. We propose a new algorithm, the tinker algorithm, to determine this core structure and apply it in the cell-cycle system. With this algorithm, the backbone network of the cell-cycle network can be determined accurately and efficiently in various models such as the Boolean model, stochastic model, and ordinary differential equation model. Results show that our algorithm is more efficient than that used in the previous research. We hope this method can be put into practical use in relevant future studies. \\ \\
\textbf{\large Keywords:} Biological network, Backbone network, Tinker algorithm, Mathematical model\\
\textbf{PACS numbers:} 07.05.Kf, 07.05.Tp

\section{Introduction}
One of the main problems in systems biology is revealing the
relationship between functions and components (proteins, nucleic acids, polysaccharides) as well as their interactions in biological networks
\cite{Kauffman1993, Alon2006, Rigoutsos2007, Pang2008}. With the contribution of
experimental physiologists, the precise details of components, interactions,
and full map of the biological regulatory network have been presented.
To analyze the function of the network, many
mathematical methods have been introduced, such as master equations model, the Monte Carlo method \cite{Gillespie1976}, the stochastic
model \cite{Jong2002,Cai2009,Zhang2006}, ordinary differential equation (ODE) model
\cite{Katherine2000, Novak2001, Tyson2001,Tyson2003,Csikasz2009}, and the Boolean network method
\cite{Kauffman1969, Bornholdt2005, Li2004, Wang2010,Xia2011,Wang2012}.
These mathematical models not only help us understand how the
biological components work, but also provide many useful and precise
predictions \cite{Davidich2008a, Davidich2013}.

An important application of the mathematical models is to investigate how the biological network is robustly designed for realizing its function. Creatures should be well adapted to the environment after the long process of evolution. Sometimes, the adaptability is expressed as the high robustness of the function and structure against perturbations \cite{Kitano2004,Stelling2004,Wagner2005}. One of the well-studied examples is the research on the cell-cycle network that describes the metabolic process \cite{Cooper2000,Morohashi2002,Lichtenberg2005}. Li {\it et al.} have investigated the global dynamical property and stability of the cell-cycle network of budding yeast in the Boolean network model \cite{Li2004} and the ODE model \cite{Li2014}, and revealed that the cell-cycle pathway is an attracting trajectory of the dynamics and rather stable against perturbations. They have also studied the robustness of the network in the stochastic model from the potential landscape perspective \cite{Zhang2006}. Wang {\it et al.} have also carried out many works from the potential landscape perspective in the study of the cell-cycle network using the stochastic model and the ODE model \cite{Wang2010, Xia2011, Wang2012}. Their results have demonstrated that the network has the least dissipation cost design for its robustness. These studies that use various models help us understand the biological network's robustness further.

The biological network is robustly designed as a whole, but the functional role of local edges would be diverse. Several works of Wang {\it et al.}, on the basis of the Boolean network model, have revealed that the functional backbone motif structures of the cell-cycle network of budding and fission yeast, where the functional backbone motif is the minimal subnetwork (with the fewest edges) that can maintain the main function of the full network \cite{Wang2010, Xia2011, Wang2012, Zhang2013}. Afterward, Yang {\it et al.} constructed a simple cell-cycle network of cancer cells and found that the backbone motif structure also exists in this network \cite{Yang2010}. From this perspective, a biological network can be viewed as the combination of two components: the functional backbone motif (backbone network), which carries out the main function of the network; and the remaining edges (supplementary edges), which can enhance the robustness of the network. This classification could not only help us understand why a network still works normally after some regulatory edges are removed, but also provide us considerable information to control the network or even design a new artificial network in further experiments. Therefore, finding the backbone network is a very important task.

The method used by Wang {\it et al.} is rather good in finding the backbone network, but it has some limitations. The method is process based: firstly, the Boolean network equations can be established and simplified with the known biological process; then, the simplified equations are solved by enumeration, and all the solutions are called the possible networks of the biological process; and finally, the backbone network is selected from the possible networks with the known biological network. However, this method can be applied only to the Boolean network model, and is not adaptable to the other models previously mentioned. Moreover, the process of solving equations by enumeration may cost considerable time in many situations. Therefore, just as in the case of studies on the robustness of the cell-cycle network, an efficient method that can be applied to determine the backbone network in different models is considered to be very necessary.

In this paper, we suggest a new method to determine the backbone network of the biological network based on the idea of tinkering. The idea of tinkering was firstly used in evolution by Fran\c{c}ois Jacob, who considered the process of evolution as a tinker's tinkering \cite{Jacob1977}. Then, Alon {\it et al.} proposed that the tinker is a good engineer for some biological networks' motifs, which are well designed to enhance the network robustness\cite{Alon2006,Alon2003,Alon2007,Alon2002}.
Here, based on the idea of tinkering, we propose a simple algorithm, the tinker algorithm, with which the backbone network of a given
network can be determined by tinkering the edges of the network one by one.
We apply it to different models such as the Boolean network model,
stochastic model, and ODE method. Simulation results show that the algorithm works satisfactorily.

The paper is organized as follows. In Sec. II, the description of the tinker
algorithm is provided. In Sec. III, the simulation results of the tinker
algorithm applied in the Boolean network model, stochastic model, and
ODE method are presented in detail. Finally, our conclusion and
discussions are given.

\section{Methods}
A biological network with $N$ interacting molecules
can be expressed as an adjacency matrix
$A=\{a_{ij},~i,j=1,2,...,N\}$, where each node denotes a molecule.
The value of $a_{ij}$ $(j\ne i)$ can be positive,
negative, or zero, depending on whether node $j$ activates,
inhibits, or does not interact with node $i$, respectively. The value of
$a_{ii}$ can also be positive, negative, or zero, representing that node $i$ has self-degradation, self-stimulation, or no
action back on itself, respectively. We use
$S_i(t)$ to denote the concentration or state of the $i$th
molecule at the moment $t$, where $S(t)=\{S_1(t), S_2(t), ..., S_N(t) \}$ for
all the molecules. Here, the function of a network is expressed as a special time series of $S(t)$, say $S^0(t)=\{S^0_1(t), S^0_2(t), ..., S^0_N(t)\}$; we call it the functional sequences. To determine the backbone network, two basic assumptions are necessary: the full network and functional sequences should be known from experiments or database in advance; and every node is indispensable in maintaining the function of the network, and cannot be removed.

\begin{figure}[!ht]
{\includegraphics[width=1.0\columnwidth]{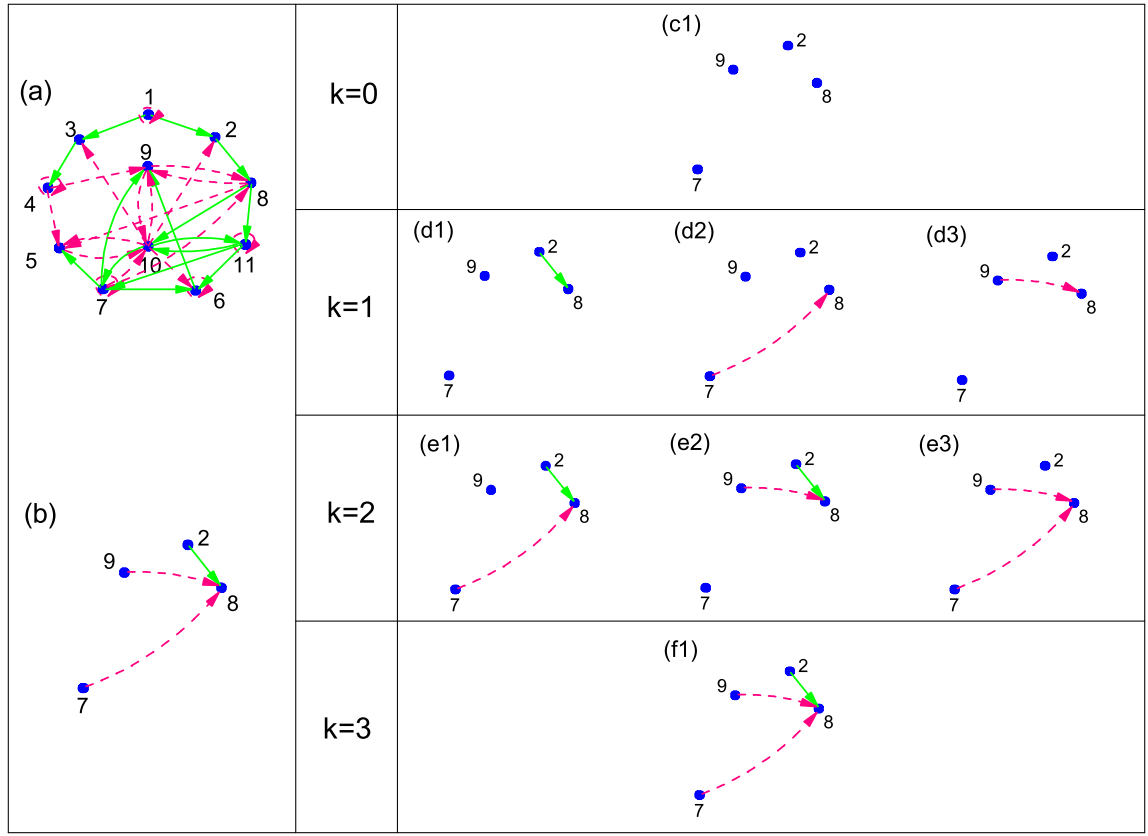}}\caption{Steps to determine the subbackbone network of a node. (a) The full network. (b) The local network of a node, for example, node ${i_0}=8$ with entry degree $d_{i_0}=3$. All the subnetworks of the local network are shown in: (c1) one case for $k=0$; (d1), (d2), and (d3)
three cases for $k=1$; (e1), (e2), and (e3) three cases for $k=2$; (f1) one case for $k=3$. Each $k$ indicates that the subnetwork of the local network contains $k$ edges.
Throughout this paper, the green arrows and pink dashed arrows
respectively represent the positive regulations and ``deactivation"
(inhibition, repression, degradation) regulations, unless otherwise
noted.}\label{fig1}
\end{figure}

The basic idea of the tinker algorithm is to identify the functional role by
tinkering each edge. We check whether the functional sequences are still maintained under the remaining network after removing an edge. If this is true, this edge is a supplementary edge and can be removed; otherwise, the edge is a backbone edge, and should be retained. For discrete models, such as the Boolean network model and the stochastic model, the function being maintained means that the new time series (after an edge is removed) is completely in accordance with the functional sequences, namely, $S(t)=S^0(t)$. Obviously, each node's functional sequence is influenced only by its neighbors. Therefore, we can construct a local network for each node: the target node and its entry regulations. In addition, there must be a minimal subnetwork of the local network under which the functional sequence of the node is maintained, and we call it the subbackbone network. Then, the backbone network of the full network can be viewed as the combination of subbackbone networks of all nodes. The schematic showing how to determine the subbackbone network of each node is in Fig.~\ref{fig1}. For the full network (Fig.~\ref{fig1}(a)), we arbitrarily select a node and construct its local network (Fig.~\ref{fig1}(b)). Then, we check its subnetworks in the order of the subnetwork containing $0,1,2,..., d_i$ edges (Fig.~\ref{fig1}(c), (d), (e), (f)) to identify which subnetwork is the subbackbone network. With this idea, the code of the tinker algorithm can be written as follows:

\definecolor{shadecolor}{rgb}{0.92,0.92,0.92}
\begin{shaded}
{ \noindent \bf // the C pseudo-code of the tinker algorithm \\
\\
tinker($A$, $I$)\{
// $I$ denotes the set of nodes.\\
$~~~$   $I_{remain}$=I;    \\
$~~~$   do( select node $i$  $\in$  $I_{remain}$  ) \{    \\
$~~~$$~~~$      // construct the local network of node $i$, $e_{ij}$ denotes the edge from $j$ to $i$.\\
$~~~$$~~~$      select $E^0_i=\{e_{ij}|a_{ij} \ne 0 , j=1,...,N\}$;    \\
$~~~$$~~~$      //find the subnetwork of $E_i^{0}$ that has the fewest entry regulations of\\
$~~~$$~~~$      //node i and under which the functional sequence can still be maintained.\\
$~~~$$~~~$      mark=false;  \\
$~~~$$~~~$      for( k=0 ; k $<=$ $d_i$ ; k++) \{    \\
$~~~$$~~~$$~~~$          for( l=1; l $<=$ $C^k_{d_i}$ ; l++) \{    \\
$~~~$$~~~$$~~~$$~~~$         // update $E_i$ with $k$ edges of $E^0_i$ retained, each $k$ corresponds to \\
$~~~$$~~~$$~~~$$~~~$         // $C^k_{d_i}$ cases, and $l$ denotes ergodicity for different cases. \\
$~~~$$~~~$$~~~$$~~~$         $E_i=update(E^0_i, k, l)$;    \\
$~~~$$~~~$$~~~$$~~~$         // the subroutine $check(S,E)$ is different for different models. \\
$~~~$$~~~$$~~~$$~~~$         if($check(S^0,E_i)$==true)mark=true, save $E_i$, break;    \\
$~~~$$~~~$$~~~$          \}    \\
$~~~$$~~~$$~~~$        if (mark==true) break; \\
$~~~$$~~~$      \}    \\
$~~~$$~~~$     remove node i from $I_{remain}$;\\
$~~~$   \}while( $I_{remain}$ != $\Phi$ );    \\
$~~~$   // obtain and check the result of the backbone subnetwork.\\
$~~~$   $E_{result}$ = $\bigcup_{i=1}^N{E_i}$; $check(S^0,E_{result})$;    \\
\}
}
\end{shaded}

Next, we analyze the computational complexity of this algorithm. For a node $i$ with entry degree $d_i$, the subnetworks of its local network can be organized as shown in Fig.~\ref{fig1}. For each case of $k$ and $l$, $k=0,1,2,.., or~d_i$ and $l=1,2,...,or~C_{d_i}^{k}$, we update the network and use the subroutine $check$ to check whether the functional sequences can be maintained under the new network. This process will cost most time in running the program since the number of cases that we have to check is
\begin{equation}\label{cases}
m_i=C_{d_i}^{0}+C_{d_i}^{1}+C_{d_i}^{2}+...+C_{d_i}^{d_i-1}+C_{d_i}^{d_i}=2^{d_i}.
\end{equation}
Actually, the program may stop in the middle of the loop (if the
check is true); thus, less than $m_i$ cases are needed to be considered in practical conditions. Moreover, for a network with degree distribution
$d_1, d_2, ..., d_N$, we consider $T_m=\sum_{i=1}^{N}m_i=\sum_{i=1}^{N}2^{d_i}$
cases at most. Since $T_m<N2^{d_{max}}$, the computational time in which we determine the backbone network of a network with $N$ nodes is less than the order of $\vartheta(N2^{d_{max}}$), where ${d_{max}}$ is the largest degree of the nodes.

We also compare the efficiency of the tinker algorithm with that of the algorithm in the previous reference. To find all the possible networks of the functional sequences, most of the time for the algorithm used in reference \cite{Wang2010} is spent in solving the Boolean network equations by enumerating the types of regulation between any two nodes. Because the
types of regulation can be activation, inhibition, or no regulation, the time needed to solve the  equations (not simplified) is of the order of $\vartheta(N3^{N})$. If the equations are simplified, the time would be less. For the tinker algorithm, the running time of finding the backbone network is determined by $d_{max}$. For many biological networks, the average degree of each network is approximately between $2$ and $4$, where $d_{max}\ll{N}$ usually \cite{Li2004,Oosawa2002,Bornholdt2000,Costanzo2001,Lee2002}. Obviously, $\vartheta(N2^{d_{max}})$ is much smaller than $\vartheta(N3^{N})$, indicating that the tinker algorithm is more efficient.

The tinker algorithm is more efficient mostly because it makes full use of the given network. Once the full network is given, in order to identify the functional role of an edge, only two cases (keeping or removing the edge) are needed to be checked. On the other hand, for the algorithm in reference \cite{Wang2010}, the full network is not required at first. To identify the type of regulation between any two nodes, three cases (activation, inhibition, or no regulation) should be checked. This leads to the computational complexity of $\vartheta(N2^{d_{max}})$ versus $\vartheta(N3^{N})$ for the tinker algorithm and the previous algorithm. However, we can do nothing with the tinker algorithm if the full network is not given in advance. Therefore, it should be noted that the tinker algorithm is preferred only for the task of finding the backbone network of a given full network. We will focus our attention on this task in the following sections.

\section{Simulation results}
In this paper, we take the cell-cycle networks as our examples \cite{Li2004}. The full process of the
cell cycle consists of four phases: G$1$, S, G$2$, and M. In the G$1$ phase,
the cell grows, and undergoes division under appropriate conditions. Then, in the S phase, DNA is synthesized and chromosomes are replicated.
The G$2$ phase is a ``gap" phase between S and M. The final phase M
corresponds to mitosis, in which chromosomes are separated and the
cell is divided into two. Eventually, after the M stage, the cell
enters G$1$, thereby completing a ``cycle." The cell-cycle networks of budding yeast and fission yeast are shown in Fig.~\ref{fig2}(a) and Fig.~\ref{fig2}(c), respectively \cite{Li2004, Davidich2008a}. The network for budding yeast consists of eleven ($N=11$) proteins: Cln$3$, MBF, SBF, Cln$1$,-$2$, Cdh$1$, Swi$5$, Cdc$20$/Cdc$14$,
Clb$5$,-$6$, Sic$1$, Clb$1$,-$2$, and Mcm$1$/SFF, whereas the network for fission yeast includes nine ($N=9$) proteins: SK, Cdc$2$/Cdc$13$,
Ste$9$, Rum$1$, Slp$1$, Cdc$2$/Cdc$13^*$, Wee$1$/Mik$1$, Cdc$25$, and PP. We use $S^0(t)=\{S^0_i(t), i=1,2,...,N\}$ to express the biological process (the functional sequences) of the cell cycle. The functional sequences $S^{0}(t)$ for these two types of yeast are shown in Table \ref{Table1}.

\begin{table}[!ht]  \scriptsize
\begin{center}
\renewcommand\arraystretch{1.2}
\caption{The time course of the cell-cycle process of budding yeast (a) and fission yeast (b). This time course is also called the functional sequences of the cell-cycle network.}
{\bf(a)}\\
\begin{tabular}{l|cccccccccccccc}
\hline
Time &Cln$3$  &MBF  &SBF  &Cln$1,2$  &Cdh1 &Swi5  &Cdc$20/14$  &Clb$5,6$  &Sic$1$ &Clb$1,2$ &Mcm$1$/SFF  \\
\hline
t &$S_1^0$ &$S_2^0$ &$S_3^0$ &$S_4^0$ &$S_5^0$ &$S_6^0$ &$S_7^0$
&$S_8^0$ &$S_9^0$ &$S_{10}^0$ &$S_{11}^0$\\
\hline
0  &1 &0 &0 &0 &1 &0 &0 &0 &1 &0 &0 \\
1  &0 &1 &1 &0 &1 &0 &0 &0 &1 &0 &0 \\
2  &0 &1 &1 &1 &1 &0 &0 &0 &1 &0 &0 \\
3  &0 &1 &1 &1 &0 &0 &0 &0 &0 &0 &0 \\
4  &0 &1 &1 &1 &0 &0 &0 &1 &0 &0 &0 \\
5  &0 &1 &1 &1 &0 &0 &0 &1 &0 &1 &1 \\
6  &0 &0 &0 &1 &0 &0 &1 &1 &0 &1 &1 \\
7  &0 &0 &0 &0 &0 &0 &1 &0 &0 &0 &1 \\
8  &0 &0 &0 &0 &1 &1 &1 &0 &1 &0 &0 \\
9  &0 &0 &0 &0 &1 &1 &0 &0 &1 &0 &0 \\
10 &0 &0 &0 &0 &1 &0 &0 &0 &1 &0 &0 \\
11 &0 &0 &0 &0 &1 &0 &0 &0 &1 &0 &0 \\
\hline \hline
 \end{tabular}

\vspace{10pt}

{\bf (b)}\\
\begin{tabular}{l|ccccccccccccccc}
\hline
Time &Sk  &Cdc$2/13$  &Ste$9$  &Rum$1$  &Slp$1$  &Cdc$2/13^*$  &Wee$1$  &Cdc$25$ &PP\\
\hline t &$S_1^0$ &$S_2^0$ &$S_3^0$ &$S_4^0$ &$S_5^0$ &$S_6^0$ &$S_7^0$ &$S_8^0$ &$S_9^0$\\
\hline
0  &1 &0 &1 &1 &0 &0 &1 &0 &0\\
1  &0 &0 &0 &0 &0 &0 &1 &0 &0\\
2  &0 &1 &0 &0 &0 &0 &1 &0 &0\\
3  &0 &1 &0 &0 &0 &0 &0 &1 &0\\
4  &0 &1 &0 &0 &0 &1 &0 &1 &0\\
5  &0 &1 &0 &0 &1 &1 &0 &1 &0\\
6  &0 &0 &0 &0 &1 &0 &0 &1 &1\\
7  &0 &0 &1 &1 &0 &0 &1 &0 &1\\
8  &0 &0 &1 &1 &0 &0 &1 &0 &0\\
9  &0 &0 &1 &1 &0 &0 &1 &0 &0\\
\hline \hline
 \end{tabular}

  \label{Table1}
  \end{center}
\end{table}

\subsection{Tinker algorithm applied in the Boolean network model}
In the Boolean network model, each node $i$ has only two states,
$S_i(t)=1$ or $0$, representing the ``on" (active) or ``off"
(inactive) state of the molecule, respectively. Therefore, there are
$2^N$ possible states of the networked system. The general dynamical rule, based on which
the molecular states in the next time step can be determined by the
molecular states in the current time step, is described as follows
\cite{Li2004}:
\begin{eqnarray}\label{Boolean}
   S_i(t+1)=\left\{\begin{array}{ccccccccc}
      1, &\qquad \text{if}~\sum_{j=1}^{N}a_{ij}S_j(t)+h_i>0 & \\
      0, &\qquad \text{if}~\sum_{j=1}^{N}a_{ij}S_j(t)+h_i<0 & \\
      S_i(t), &\qquad \text{if}~\sum_{j=1}^{N}a_{ij}S_j(t)+h_i=0,
      \end{array}\right.
\end{eqnarray}
where $h_i$ is the threshold. Here, $h_i=0$, $a_{ii}$ take the value $-1$, $1$, or $0$, and $a_{ij}$ take the value
$-\gamma$, $1$, or $0$ for $j\ne i$. In the dominant inhibition model,
where inhibition is dominant over stimulation for other types of
interactions, namely, $\gamma=\infty$, Eq. (\ref{Boolean}) can
be simplified into a logical equation \cite{Wang2010}
\begin{equation}\label{Booleans}
S_i(t+1)=(\sum_{j\ne i}(S_j(t)g_{ij})+S_i(t)
\overline{r}_{ii}+\overline{S_i(t)}g_{ii})\prod_{j\ne
i}(\overline{S_j(t)r_{ij}}),
\end{equation}
where $r_{ij}$ represents a putative inhibitory edge and $g_{ij}$
stands for a putative stimulatory edge from node $j$ to $i$.
The addition, multiplication, and bar in the equation represent the Boolean
operators OR, AND, and NOT, respectively. In this model, the
subroutine $check$ is described as follows:

\definecolor{shadecolor}{rgb}{0.92,0.92,0.92}
\begin{shaded}
{
\noindent \bf // the  subroutine $check$ for the Boolean model \\
\bf check($S^0$,$E_i$)\{              \\
$~~~$     $l_s$=length of $S^0(t)$;             \\
$~~~$      flag=true;             \\
$~~~$      for(t=0;t$<$ $l_s$;t++)\{             \\
$~~~$$~~~$          $S(t)$=$S^0(t)$;             \\
$~~~$$~~~$          $S_i(t+1)$=$eq(S(t),i)$; // here, $eq(~)$ is equation (\ref{Boolean}) or (\ref{Booleans})             \\
$~~~$$~~~$          if( $S^0_i(t+1)$ != $S_i(t+1)$) flag=false, break;         \\
$~~~$      \}              \\
$~~~$     return flag;             \\
\}             \\
}
\end{shaded}

\begin{figure}[!ht]
{\includegraphics[width=1.0\columnwidth]{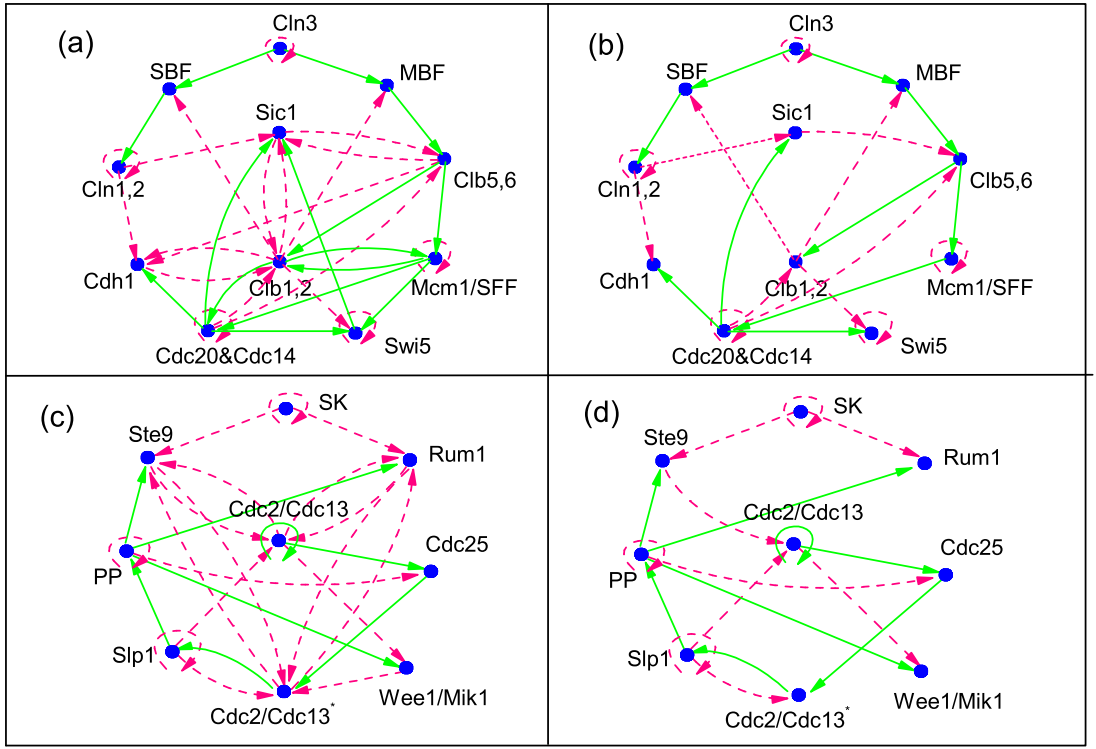}}\caption{(a) and (c) The full
cell-cycle networks of budding yeast and fission yeast, respectively. (b) and (d) The backbone networks of budding yeast and fission yeast in the Boolean network model, respectively. There are $34$ edges for the full cell-cycle network of budding yeast, while only $23$ edges are retained in its
backbone network. There are $26$ edges in the full cell-cycle network
of fission yeast, while only $18$ edges are retained in its backbone network. The backbone networks in the stochastic model are the same as those in the Boolean network model.}\label{fig2}
\end{figure}

Based on the tinker algorithm and the subroutine of the dominant inhibition Boolean
network model, the backbone networks of the cell-cycle networks of budding and fission yeast are determined and shown in Fig.~\ref{fig2}(b) and Fig.~\ref{fig2}(d), respectively. Through careful comparison, we find that they are identical to those shown in reference \cite{Wang2010}. This verifies that the tinker algorithm is effective.

\begin{figure}[!ht]
{\includegraphics[width=1.0\columnwidth]{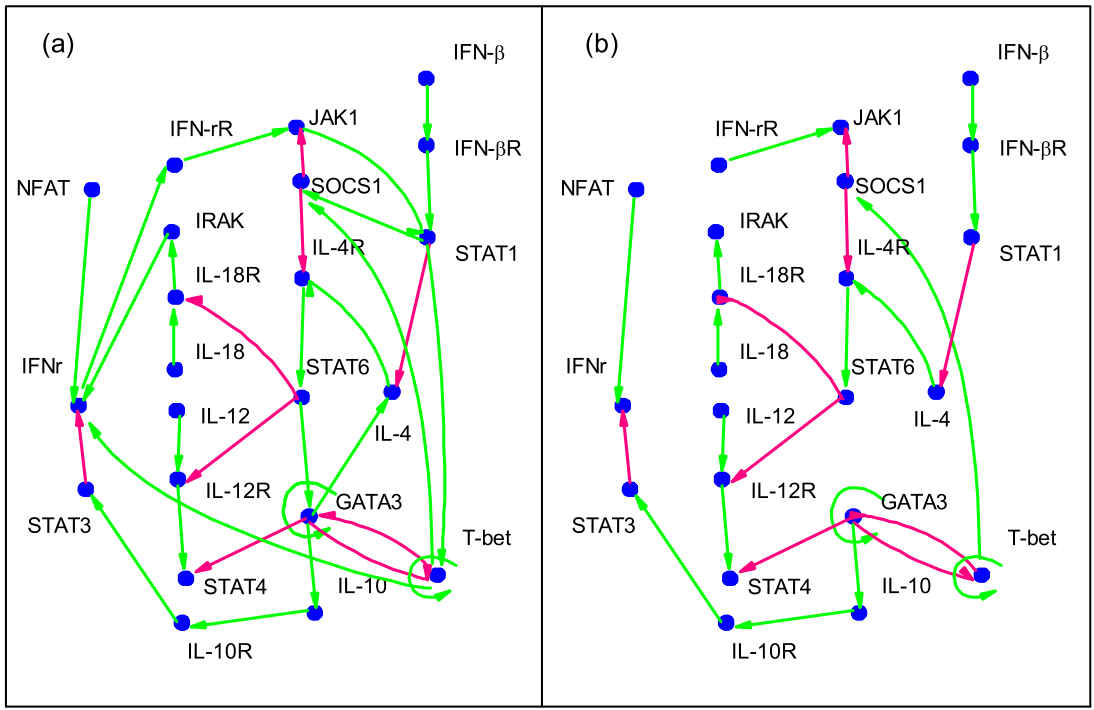}}\caption{(a) The regulatory network that controls the differentiation process of T helper cell. It is almost identical to the network in reference \cite{Mendoza2006} except that the node $TCR$ is removed here. (b) The backbone network of the T helper cell network. There are $34$ edges in the full T-helper cell network, while only $24$ edges are retained in its backbone network.}\label{fig3}
\end{figure}

Furthermore, we apply the tinker algorithm to find the backbone network of a larger network. As shown in Fig.~\ref{fig3}(a), the regulatory network that controls the differentiation process of the T helper cell contains $22$ nodes: IL-$18$, IL-$12$, IFNr, IL-$4$, IL-$10$, NFAT, IL-$18$R, IL-$12$R, IFN-rR, IL-$4$R, IL-$10$R, IFN-$\beta$, IRAK, STAT$4$, JAK$1$, STAT$6$, STAT$3$, IFN-$\beta$R, SOCS$1$, T-bet, STAT$1$, and GATA$3$ \cite{Mendoza2006}. In addition, the time course shown in Table \ref{Table2} is chosen to be the functional sequences ($S^{0}(t)$) of the nodes. As one can see, the largest degree of the nodes is $7$ ($d_{max}=7$), which is much smaller than the network size. According to the theoretical result in Section II, very short time would be enough to determine the backbone network with the tinker algorithm. In fact, our simulation result shows that we can obtain the backbone network (Fig.~\ref{fig3}(b)) of the regulatory network in $10^{-4}$ seconds. This demonstrates that the tinker algorithm is really efficient.

\begin{table}[!ht]
\scriptsize
\begin{center}
\renewcommand\arraystretch{1.0}
\caption{The time course of the T-helper cell network.~~~~~~~~~~~~~~~~~~~~~~~~~~~~~~~~~}
\begin{tabular}{l|p{0.25cm}p{0.25cm}p{0.25cm}p{0.25cm}p{0.25cm}p{0.25cm}p{0.25cm}p{0.25cm}p{0.25cm}p{0.25cm}p{0.25cm}
p{0.25cm}p{0.25cm}p{0.25cm}p{0.25cm}p{0.25cm}p{0.25cm}p{0.25cm}p{0.25cm}p{0.25cm}p{0.25cm}p{0.25cm}}
\hline
Time &$S_1^0$ &$S_2^0$ &$S_3^0$ &$S_4^0$ &$S_5^0$ &$S_6^0$ &$S_7^0$
&$S_8^0$ &$S_9^0$ &$S_{10}^0$ &$S_{11}^0$ &$S_{12}^0$ &$S_{13}^0$ &$S_{14}^0$ &$S_{15}^0$ &$S_{16}^0$ &$S_{17}^0$ &$S_{18}^0$ &$S_{19}^0$ &$S_{20}^0$ &$S_{21}^0$ &$S_{22}^0$\\
\hline
0 &1	&1	&0	&1	&0	&1	&0	&0	&1	&0	&0	&1	&0	&0	&0	&0	&0	&0	&0	&0	&0	&0\\
1 &1	&1	&1	&1	&0	&1	&1	&1	&1	&1	&0	&1	&0	&0	&1	&0	&0	&1	&0	&1	&0	&1\\
2 &1	&1	&1	&1	&1	&1	&1	&1	&1	&1	&0	&1	&1	&0	&1	&1	&0	&1	&1	&0	&1	&0\\
3 &1	&1	&1	&0	&1	&1	&0	&0	&1	&0	&1	&1	&1	&1	&0	&1	&0	&1	&1	&1	&1	&1\\
4 &1	&1	&1	&0	&1	&1	&0	&0	&1	&0	&1	&1	&1	&0	&0	&1	&1	&1	&1	&0	&1	&0\\
5 &1	&1	&0	&0	&1	&1	&0	&0	&1	&0	&1	&1	&1	&0	&0	&1	&1	&1	&1	&1	&1	&1\\
6 &1	&1	&0	&0	&1	&1	&0	&0	&1	&0	&1	&1	&1	&0	&0	&1	&1	&1	&1	&0	&1	&0\\
7 &1	&1	&0	&0	&1	&1	&0	&0	&1	&0	&1	&1	&1	&0	&0	&1	&1	&1	&1	&1	&1	&1\\
\hline \hline
 \end{tabular}

   \label{Table2}
  \end{center}
\end{table}

Although the backbone network of a full network can be found accurately with the tinker algorithm, it should be mentioned that the obtained backbone network may not be unique because of the existence of structure symmetry. From Fig.~\ref{fig2}(c) and Table \ref{Table1}(b), one can see that the role of Ste$9$ and Rum$1$ in the cell-cycle network of fission yeast share high similarities. In Fig.~\ref{fig2}(d), if the edge from Ste$9$ to Cdc$2$/Cdc$13$ is substituted by the edge from Rum$1$ to Cdc$2$/Cdc$13$, the new network is also a backbone network. As our algorithm is based on determining the subbackbone network of each node just according to the main function sequence, structure symmetry is a global property and it may not be expressed. For the network with high structure symmetry, it is not easy to find all the backbone networks for it may greatly increase the computational complexity. Hence, we do not expand this topic further.

\subsection{Tinker algorithm applied in the stochastic model}
In the stochastic model, which is based on the Boolean network model, it is assumed that the transition of the states of the molecules is a stochastic process with the Markov property \cite{Zhang2006}.
In the Boolean network model, the next state S(t + 1) from the present state S(t) is exactly determined. However, in
the stochastic model, the next state $S(t+1)$ can be an arbitrary state of the
$2^N$ states, but with a different transition probability to each state. The transition
probability is described as follows:
\begin{equation}\label{stmodel}
Pr(S(t+1)|S(t))=\prod_{i=1}^{N}Pr(S_i(t+1)|S(t)),
\end{equation}
where $Pr(S_i(t+1)|S(t))$ is the conditional probability that the
state of the $i$th node is $S_i(t+1)$ at moment $t+1$, and it is
expressed as
\begin{equation}
Pr(S_i(t+1)=\sigma_i|S(t))=\frac{e^{(2\sigma_i-1)\beta T}}{e^{\beta
T}+e^{-\beta T}},
\end{equation}
if $T=\sum_{j=1}^{N}a_{ij}S_j(t)\ne 0$, $\sigma_i\in\{0,1\}$; if
$T=\sum_{j=1}^{N}a_{ij}S_j(t)=0$, then
\begin{eqnarray}
Pr(S_i(t+1)&=&S_i(t)|S(t))=\frac{1}{1+e^{-\alpha}},
 ~~if~a_{ii}\ne 1, \\
Pr(S_i(t+1)&=&1|S(t))=\frac{1}{1+e^{-\alpha}},~~~~~~~~
 if~a_{ii}=1.
\end{eqnarray}
The values of $a_{ii}$ and $a_{ij}$ are identical to those used in
the Boolean network model. $\alpha$ and $\beta$ are two parameters, whose
robustness has been analyzed in reference
\cite{Zhang2006}. Here, we
set $\alpha=5$, $\beta=6$, and $\gamma=10$.

Next, we will show the details of the subroutine $check$ in this model. Although the next state $S(t+1)$ can be an arbitrary state of the $2^N$ states, there is a state $S^{L}(t+1)$ such that
\begin{equation}
Pr(S^{L}(t+1)|S(t))=max\{Pr(S_1(t+1),...,S_N(t+1)|S(t)), S_i(t+1)\in\{0, 1\}\},
\end{equation}
and we call it the most probable transition
state. Naturally, the state $S_i^{L}(t+1)$ for each $i$ satisfies
$Pr(S_i^{L}(t+1)|S(t))=max\{Pr(S_i(t+1)|S(t), S_i(t+1)=0,1\}$, namely,
\begin{equation}\label{LPr}
S_i^{L}(t+1)=\{S_i(t+1)|Pr(S_i(t+1)|S(t))=max\{Pr(S_i(t+1)|S(t)\}\}.
\end{equation}
In this model, the function of the network being maintained means that $S^{L}=S^{0}$.
Then, the subroutine $check$ is described as follows:
\definecolor{shadecolor}{rgb}{0.92,0.92,0.92}
\begin{shaded}
{
\noindent \bf // the  subroutine $check$ for the stochastic model  \\
\bf check($S^0$,$E_i$)\{             \\
$~~~$      $l_s$=length of $S^0(t)$;             \\
$~~~$      flag=true;             \\
$~~~$      for(t=0;t$<$ $l_s$;t++)\{             \\
$~~~$ $~~~$         $S(t)$=$S^0(t)$;             \\
$~~~$ $~~~$         $S^{L}_i(t+1)$=pr-eq($S(t)$, $i$); // here, pr-eq(~) is equation (\ref{LPr})            \\
$~~~$ $~~~$         if( $S^{L}_i(t+1)$ !=$ S^0_i(t+1)$ ) flag=false, break;             \\
$~~~$      \}             \\
$~~~$     return flag;             \\
\}
}
\end{shaded}

We still take the cell-cycle network of budding and fission yeast as
examples. Their backbone networks can be determined with this algorithm, and it turns out that they are the same as
those found in the dominant Boolean network model (Figs.~\ref{fig2}(b) and (d)).

\begin{figure}[!ht]
{\includegraphics[width=1.0\columnwidth]{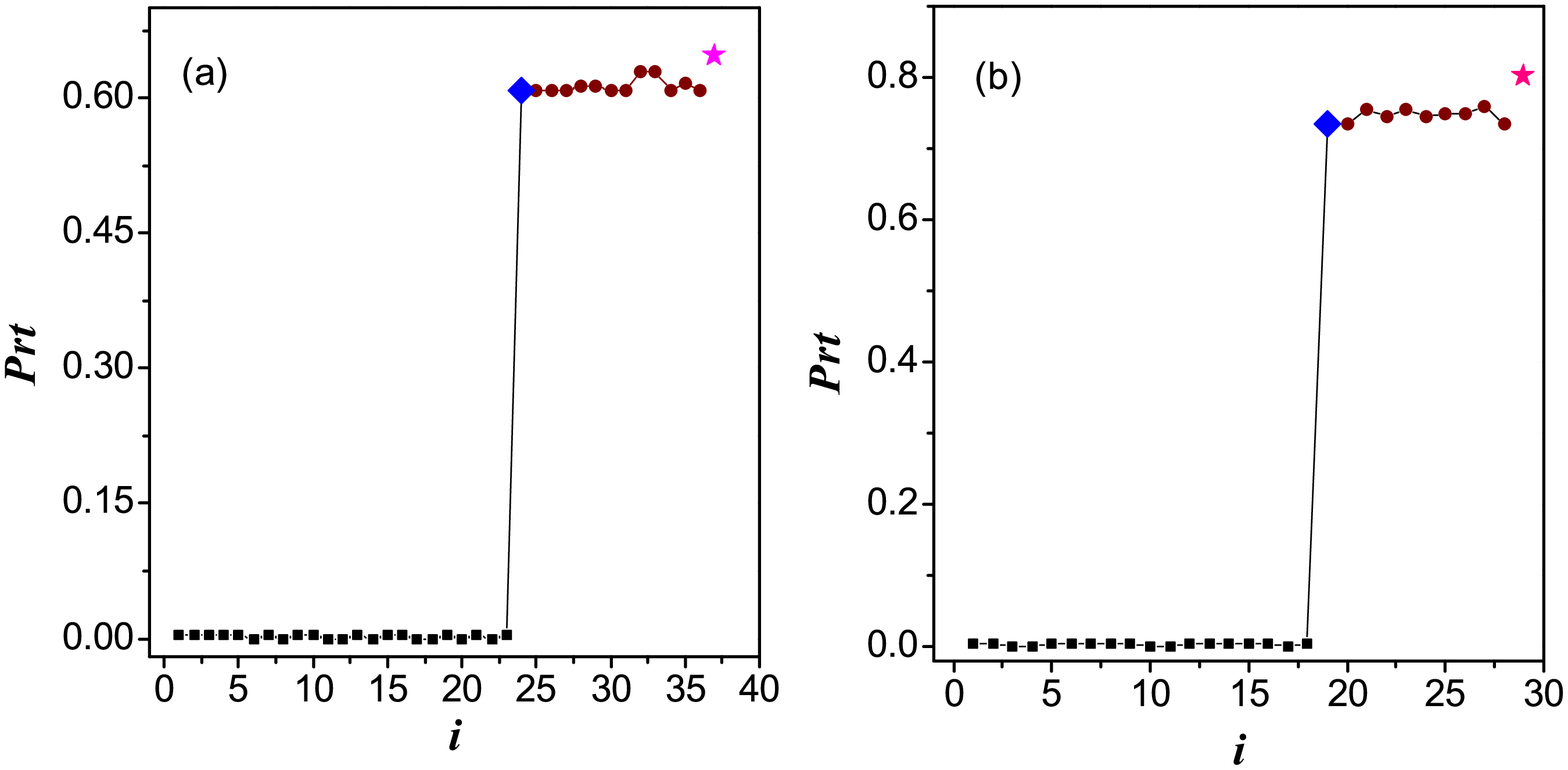}}\caption{(a) The trajectory probability $Prt$ in Eq.
(\ref{trjP}) for the backbone network with one edge removed
(first $23$ squares), for the backbone network (blue diamond), for the backbone network
with one of the supplementary edges added ($11$ dots), and for the full cell-cycle network
(pink star) of budding yeast. (b) The same as (a) for fission yeast.}\label{fig4}
\end{figure}

We also compare the importance of backbone network edges with supplementary edges (the edges
belonging to the full network but not in the backbone network). For the budding yeast, there are $34$ edges in the full network and
$23$ edges in the backbone network. We define the trajectory transition
probability as
\begin{equation}\label{trjP}
Prt=\prod_{t=0}^{t_{end}}Pr\{S^{0}(t+1)|S^{0}(t)\}.
\end{equation}
$Prt$ is calculated under the following networks: the backbone network with one edge removed ($23$ cases), the backbone network,
the backbone network with one of the supplementary edges added ($11$ cases), and
the full network (Fig.~\ref{fig4}(a)). The probability $Prt$ is nearly zero for the backbone
network with one edge removed, indicating that the network can barely carry
out the function if even one edge is lost. For the other cases, $Prt$ is much
larger, and it is the largest for the full network case. It demonstrates that the
supplementary edges really can enhance the robustness of the network.
We carry out the same process in the cell-cycle network of fission yeast
and find similar phenomena (Fig.~\ref{fig4}(b)).

\subsection{Tinker algorithm applied in the ordinary differential equation (ODE) model}
The ordinary differential equation (ODE) model is a time continuous model. It is one of the most useful
and effective models, as it can exactly give the concentration of each
molecule, which can be compared with experimental results.
Therefore, we expect to determine the functional backbone network in the ODE
model with the tinker algorithm.

Compared to the previous two models, determining the backbone network using the tinker algorithm is slightly different in the ODE model.
In the Boolean network model, to determine whether the function is maintained, we check whether the new $S(t)$ is equal to $S^{0}(t)$ after removing some edges of the network. If we still apply this standard to the ODE model, the results of the check will always be false. Because for the ODE model, the functional sequence ($S^{0}(t)$) is the continuous solution of the ODE equations under the full network.  If an arbitrary edge of the full network is removed, the new solution $S(t)$ under the new network would  always be different from $S^{0}(t)$. However, a very small difference between $S(t)$ and $S^{0}(t)$ is actually inconsequential for the function. Thus, the original standard is not appropriate for the ODE model. Therefore, we give a more reasonable standard: ($1$) the Pearson correlation coefficient ($R_i$) between $S_i^{0}(t)$ and the new $S_i(t-t_0)$ is larger than $R_0$, namely, $R_i>R_0$, for $i=1,2,...,N$; ($2$) the difference between the period of $S_i^{0}(t)$ and $S_i(t)$ is less than $T_0$, namely, $|T_i^0-T_i|<T_0$, for $i=1,2...,N$, where $S_i(t-t_0)$ is a proper time translation of $S_i(t)$. If the new standard is achieved with the proper thresholds $R_0$ and $T_0$, the basic trends of $S(t)$ and $S^{0}(t)$ would be identical except for small deviations. Then we can say that the function of the network is maintained. Under the new standard, the network should be treated as a whole and each node cannot be treated locally anymore, necessitating some modifications of the tinker algorithm. The details of the tinker algorithm, as well as the subroutine $check$, for the ODE model can be seen in Appendix A.

\begin{figure}[!ht]
{\includegraphics[width=1.0\columnwidth]{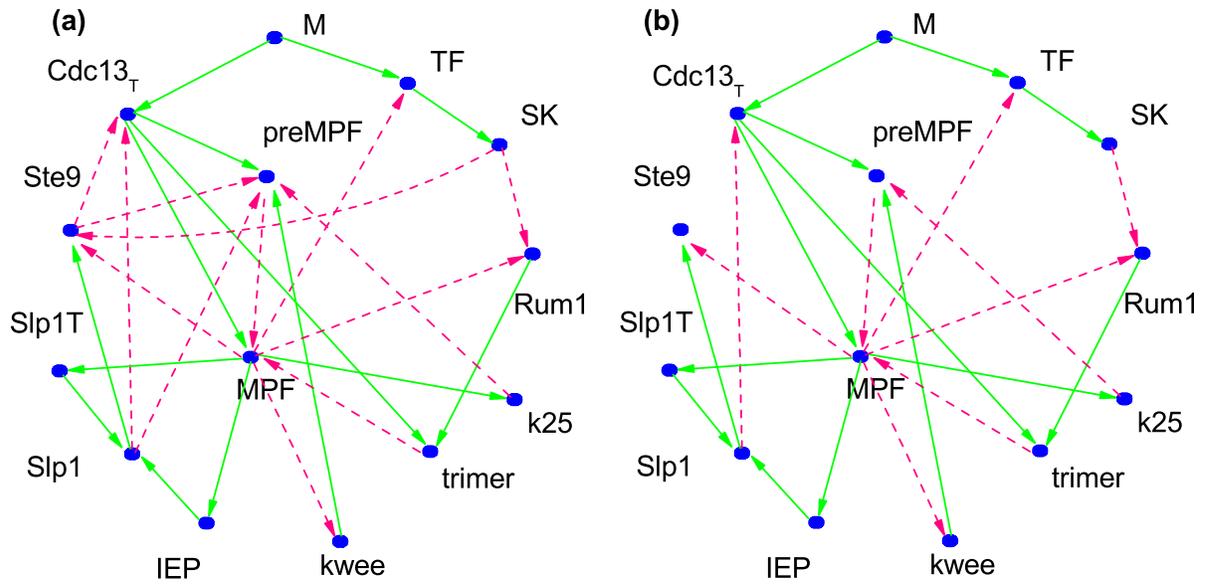}}\caption{(a) and (b) The full
cell-cycle network and the backbone network of fission yeast in the ODE model, respectively.}\label{fig5}
\end{figure}

Here, we take the cell-cycle network of fission yeast as an example to show the results. Many ODE models for the cell-cycle network of fission yeast have been introduced \cite{Novak1997, Tyson2001, Sveiczer2000, Li2010}; here we use the model in references \cite{Novak2001} and \cite{Davidich2008b} as it can well explain the variation in concentration of the proteins during the process of the cell cycle. The network for this ODE model is shown in Fig.~\ref{fig5}(a). We use $S_1(t), S_2(t),..., S_{14}(t)$ to represent the concentrations or values of Cdc$13_{T1}$, preMPF$_{1}$, Ste$9_1$, Slp$1_{T1}$, Slp$1_1$, IEP$_1$, Rum$1_{T1}$, SK$_1$, M, TF$_1$, k$_{wee_1}$, k$_{25_1}$, MPF$_1$, and Trimer$_1$, respectively; the subscript $1$ marks the rescaled variables with maximum $1$. The equations for $S(t)$ are described as follows:
\begin{eqnarray}
\notag \frac{dS_1}{dt}&=&a_{1,9}k_1S_9-(k_2^{'}+a_{1,3}k_2^{''}S_3+a_{1,5}k_2^{'''}S_5)S_1,\\
\notag \frac{dS_2}{dt}&=&a_{2,11}k_0^{'}S_{11}(a_{2,1}k_0^{''}S_1-S_2)-a_{2,12}k_0^{'''}S_{12}S_{2}-(k_2^{'}+a_{2,3}k_2^{''}S_3+a_{2,5}k_2^{'''}S_5)S_2,\\
\notag \frac{dS_3}{dt}&=&a_{3,5}(k_3^{'}+k_3^{''}S_5)\frac{1-S_3}{J_3+1-S_3}-(a_{3,8}k_4^{'}S_8+a_{3,13}k_4S_{13})\frac{S_3}{J_4+S_3}  ,\\
\notag \frac{dS_4}{dt}&=&k_5^{'}+a_{4,13}k_5^{''}\frac{S_{13}^4}{J_5^{4}+S_{13}^4}-k_6S_4  ,\\
\notag \frac{dS_5}{dt}&=&a_{5,6}k_7S_6\frac{a_{5,4}(S_4-S_5)}{J_7+S_4-S_5}-k_8\frac{S_5}{J_8+S_5}-k_6S_5  ,\\
\notag \frac{dS_6}{dt}&=&a_{6,13}k_9S_{13}\frac{1-k_9^{'}S_6}{J_9+1-k_9^{'}S_6}-k_{10}\frac{k_9^{'}S_6}{J_{10}+k_9^{'}S_6}  ,\\
\notag \frac{dS_7}{dt}&=&k_{11}-(k_{12}+a_{7,8}k_{12}^{'}S_8+a_{7,13}k_{12}^{''}S_{13})S_7  ,\\
\notag \frac{dS_8}{dt}&=&a_{8,10}k_{13}S_{10}-k_{14}S_8  ,\\
\notag \frac{dS_9}{dt}&=&\mu S_9,\\
\notag S_{10}&=&G(a_{10,9}k_{15}S_9, k_{16}^{'}+a_{10,13}k_{16}^{''}S_{13},J_{15},J_{16})  ,\\
\notag S_{11}&=&k_{wee}^{'}+(k_{wee}^{''}-k_{wee}^{'})G(V_{awee},a_{11,13}V_{iwee}S_{13},J_{awee},J_{iwee}),\\
\notag S_{12}&=&k_{25}^{'}+(k_{25}^{''}-k_{25}^{'})G(a_{12,13}V_{a25}S_{13},V_{i25},J_{a25},J_{i25}),\\
\notag S_{13}&=&\frac{(a_{13,1}k_{17}S_1-a_{13,2}k_{17}^{'}S_2)(a_{13,1}k_{17}S_1-a_{13,14}k_{17}^{''}S_{14})}{k_{17}^{'''}S_1},\\
\notag S_{14}&=&\frac{a_{14,1}a_{14,7}k_{18}S_1S_7}{\sigma+\sqrt{\sigma^2-k_{18}^{'}S_1S_7}},
\end{eqnarray}
where
\begin{eqnarray}
\notag \sigma &=&k_{19}^{'}S_1+k_{19}^{''}S_7+K_{diss},\\
\notag G(a,b,c,d)&=&\frac{2ad}{b-a+bc+ad+\sqrt{(b-a+bc+ad)^2-4ad(b-a)}}.
\end{eqnarray}
$a_{i,j}$ is the matrix element of the regulatory network
(Fig.~\ref{fig5}(a)), $A=\{a_{i,j}, i,j=1,...,N\}$, takes the value
$0$ or $1$ and denotes the presence or absence of interaction from
protein $j$ to $i$. The values of the parameters in the equations can be seen in Table \ref{Table3}. Moreover, the cell mass $M$ represented by $S_9$, increases exponentially before
cell division. We divide the cell mass by two ($S_9=S_9/2$) at the end
of mitosis; the time $S_{13}$ (MPF) decreases through $0.1$ when a cell splits itself into two daughter cells.

\begin{table}[!ht]  \footnotesize
\begin{center}
\renewcommand\arraystretch{1.5}
\setcounter{table}{2}
\renewcommand{\thetable}{\arabic{table}}%
\renewcommand{\thetable}{\arabic{table}}
\caption{Parameter values for the ODE model of the cell-cycle network of fission yeast.
They are almost identical to those in reference \cite{Davidich2008b} except for some
modifications.}
\begin{tabular}{l|ccccccccccccccc}
\hline \hline
S$_1$ &$k_1=0.04$, &$k_2^{'}=0.03$, &$k_2^{''}$=1.0, &$k_2^{'''}=0.21$\\
S$_2$ &$k_0^{'}=1.0$, &$k_0^{''}=1.17$, &$k_0^{'''}=5.0$\\
S$_3$ &$k_3^{'}=1.0$, &$k_3^{''}=21.0$, &$J_3=0.01$, &$k_4=50.75$, &$k_4^{'}=1.98$, &$J_4=0.01$\\
S$_4$ &$k_5^{'}=0.002$, &$k_5^{''}=0.143$, &$J_5=0.20689$, &$k_6=0.1$\\
S$_5$ &$k_7=0.429$, &$J_7=0.0005$, &$k_8=0.119$, &$J_8=0.0005$\\
S$_6$ &$k_9=0.16$, &$k_9^{'}=0.91$, &$J_9=0.01$, &$k_{10}=0.01$, &$J_{10}=0.01$\\
S$_7$ &$k_{11}=0.698$, &$k_{12}=0.01$, &$k_{12}^{'}=0.99$, &$k_{12}^{''}=4.35$\\
S$_8$ &$k_{13}=0.1$, &$k_{14}=0.1$\\
S$_9$ &$\mu=0.005$\\
S$_{10}$ &$k_{15}=3.0$, &$k_{16}^{'}=1.0$, &$k_{16}^{''}=2.9$, &$J_{15}=0.01$, &$J_{16}=0.01$\\
S$_{11}$ &$k_{wee}^{'}=0.115$, &$k_{wee}^{''}=1.0$, &$V_{awee}=0.25$, &$V_{iwee}=1.45$, &$J_{awee}=0.01$, &$J_{iwee}=0.01$\\
S$_{12}$ &$k_{25}^{'}=0.01$, &$k_{25}^{''}=1.0$, &$V_{a25}=1.45$, &$V_{i25}=0.25$, &$J_{a25}=0.01$, &$J_{i25}=0.01$\\
S$_{13}$ &$k_{17}=1.5$, &$k_{17}^{'}=1.3$, &$k_{17}^{''}=1.0$, &$k_{17}^{'''}=2.175$\\
S$_{14}$ &$k_{18}=0.441$, &$k_{18}^{'}=0.882$\\
$\sigma$ &$k_{19}^{'}=1.5$, &$k_{19}^{''}=0.147$, &$K_{diss}=0.001$\\
\hline \hline
 \end{tabular}
  \label{Table3}
  \end{center}
\end{table}

\begin{figure}[!ht]
{\includegraphics[width=1.0\columnwidth]{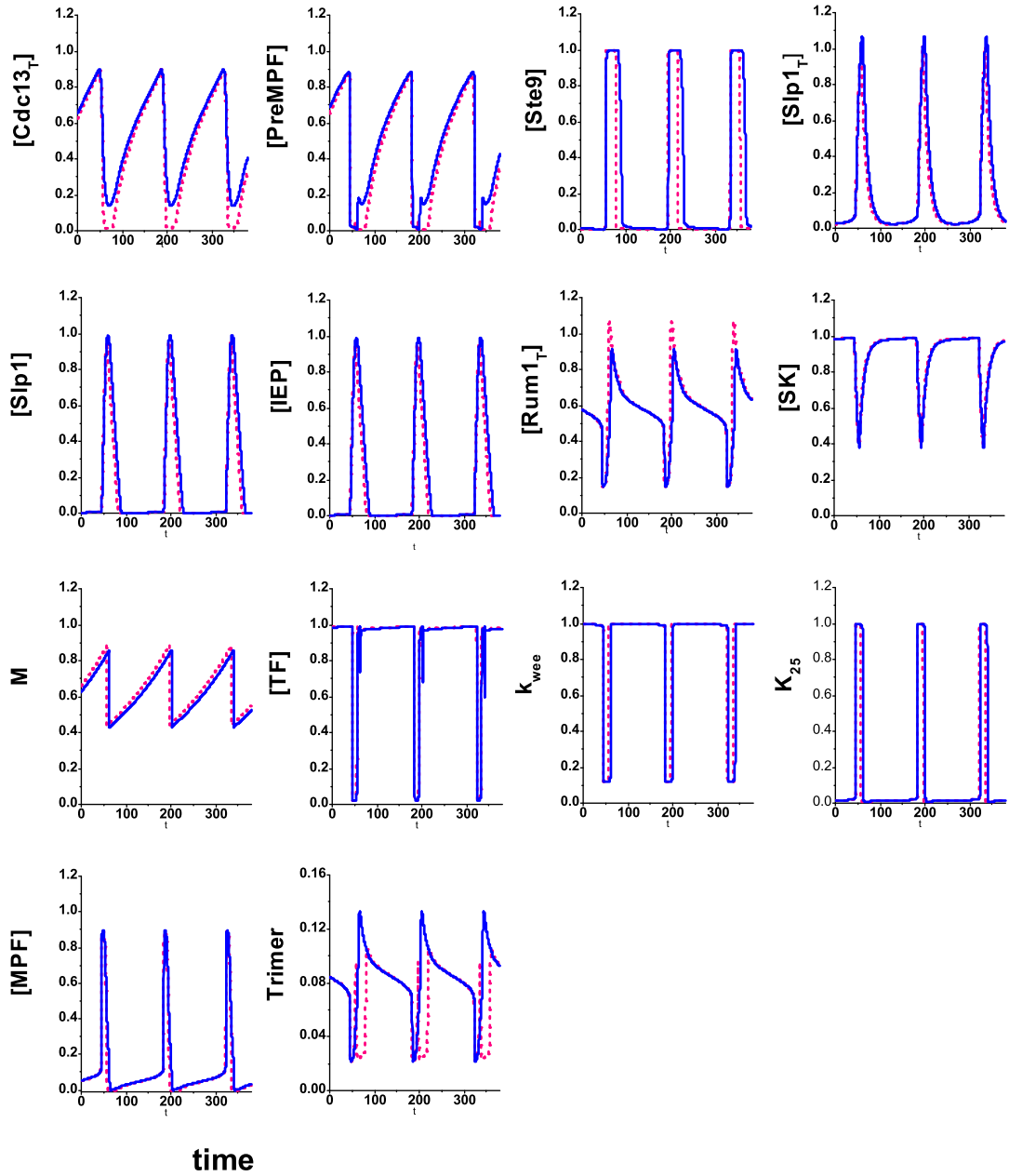}}\caption{Time
series of the solutions of the ODE model under the full network in
Fig.~\ref{fig5}(a) (pink dashed line) and the backbone network in
Fig.~\ref{fig5}(b) (blue solid line)).}\label{fig6}
\end{figure}

Based on the modified tinker algorithm and the new standard, the backbone network can be found. During our simulation, the functional sequence of the network, $S^{0}(t)=\{S_1^{0}(t),...,S_{14}^{0}(t) \}$, is defined as the solution of the ODE equations under the full cell-cycle network. Since the values of $R_0$ and $T_0$ should not be too critical or too weak, we set $R_0=0.8$ and $T_0=2\ll T_i^0$, where $T_i^0\approx138$. Accordingly, we determine the backbone network and show it in Fig.~\ref{fig5}(b). The solutions of the ODE equations under the full network and the backbone network are compared in Fig.~\ref{fig6}. One can see that they are well matched, indicating that the backbone network can really maintain the function of the full cell-cycle network, further demonstrating that the tinker algorithm is adoptable to determine the backbone network in the ODE model.

\section{Conclusion and discussions}
We have proposed a new algorithm, the tinker algorithm, with which the backbone network can be determined accurately and efficiently in
different models. In the dominant inhibition Boolean network model, the
functional backbone networks determined in the cell-cycle networks of the two yeasts match well the
results in previous works, verifying that the tinker
algorithm is effective. In the stochastic model, the functional
backbone networks could also be determined easily. At the same time, the
importance of the backbone network edges and the
supplementary edges has been tested. In addition, we found that the network loses its function if an arbitrary backbone network edge is removed, while
the supplementary edges enhance the robustness of the network. Furthermore, this method has also been applied to the ODE model, and the backbone network of the cell-cycle network of fission yeast was determined.

The tinker algorithm is quite different from the method used in
previous research papers. In reference \cite{Wang2010}, the method is mainly
based on the idea of reverse engineering and makes full use of the
message of the biological process; the goal is to determine all possible
networks just from the biological process according to the basic rules of
the dominant inhibition Boolean network model. Nevertheless, the requirement
of basic rules restricts it to only determine the backbone networks in the dominant inhibition Boolean network model.
In contrast, the tinker algorithm is based on the idea of tinkering and
makes use of the messages of both the network structure and the biological process.
The purpose is to remove the supplementary edges that have
little impact on the biological process. This method is more general and can be applied in various
mathematical models as shown in our simulation results.

\begin{figure}[!ht]
{\includegraphics[width=1.0\columnwidth]{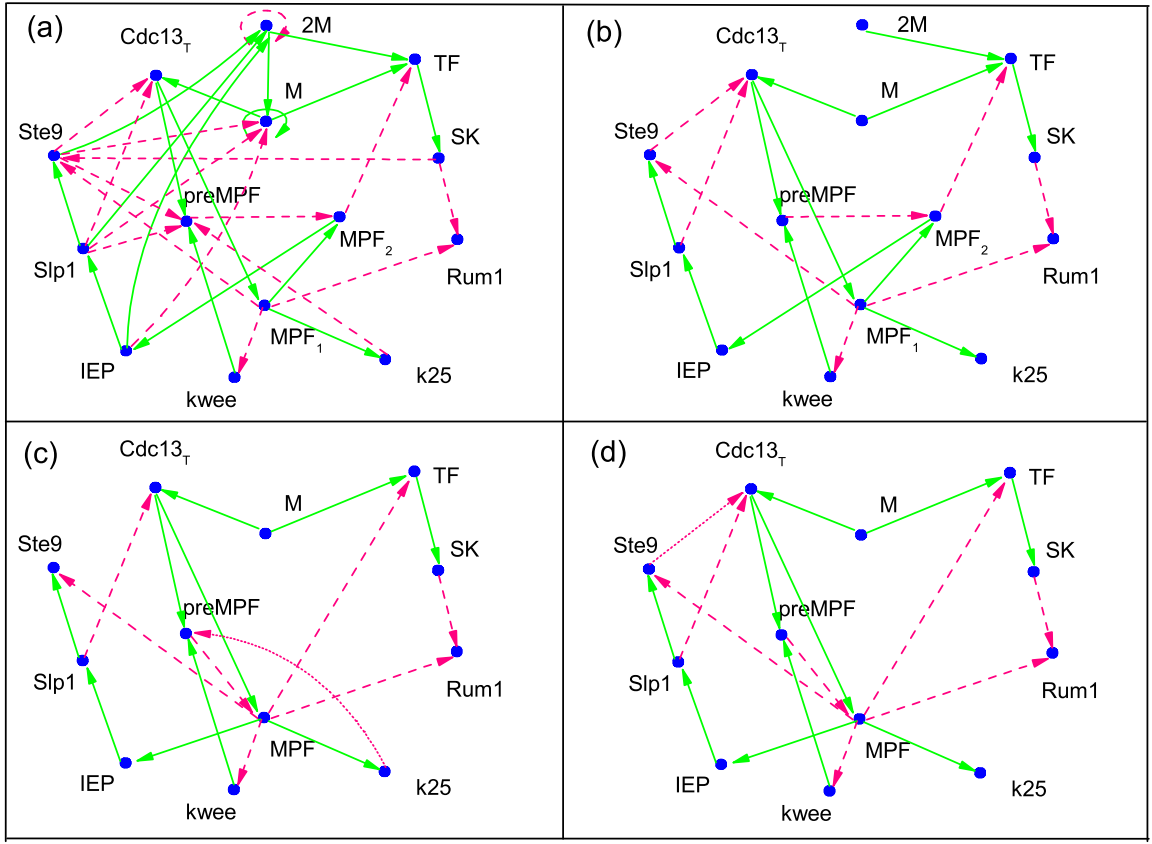}}\caption{(a) The
full network of fission yeast obtained according to the
transition from the ODE model to the Boolean network model. (b) The
backbone network of (a) in the Boolean network model.
(c) The simplified network of the backbone network
 in the ODE model (Fig.~\ref{fig5}(b)), where the nodes Slp$1T$ and
Trimer are removed. (d) The simplified network of (b), where the
node $2$M is removed, and MPF$1$ and MPF$2$ are merged. The
difference between networks in (c) and (d) is marked with dotted arrows.
}\label{fig7}
\end{figure}

With this advantage, the backbone networks obtained from different models can be compared with each other. From previous results, we know that the backbone networks in the dominant inhibition Boolean network model and the stochastic model are identical. Next, we discuss the difference between the backbone networks in the Boolean network model and the ODE model. In the previous section, the backbone network of the fission yeast cell-cycle network has been found in the ODE model. As a special coarse-grained limit of this ODE model, a Boolean model can be formulated \cite{Davidich2008b}, whose details can be seen in Appendix B. We determine the backbone network in this Boolean model and compare it with the backbone network found in the ODE model (Fig.~\ref{fig7}). Interestingly, we find that they are almost identical except that one edge is mismatched: the regulation from
k$25$ to preMPF in the ODE model is replaced by the regulation from
Ste$9$ to Cdc$13_T$ in the Boolean network model (Fig.~\ref{fig7}(c) and (d)). By analyzing the
interaction coefficients and time series associated with these four
proteins, we find that the mismatch may be attributed to the following fact: the regulations between any two proteins can be strong or weak in the ODE model, while they are all the same in the Boolean model. This indicates that the interaction strength affects the importance of the interaction (edge), and hence influences the backbone network structure.

Our aim of proposing this algorithm is to find the most important interactions in the complex biological processes. It is easy to find that the perturbation on the backbone structure rather than the supplementary edges can make a much more important effect on the dynamic process. As the evolutionary process goes on, the mutations on the supplementary structure may cause an effective process and help the creatures to adapt to the change of environment, whereas the mutation on the backbone structure usually causes the process to be terminated and adversely affects the survival and evolution. The tinker algorithm helps us distinguish the backbone network and supplementary edges, and it may give us a preliminary conclusion about the most important interactions of the network. In addition to determining the backbone network, we notice that many researchers are working to detect the core or the skeleton network of a complex network, which is defined as a network that has only a few nodes and its dynamic trajectory qualitatively approaches the full network \cite{Novak1998,Zhang2014,Liao2011}. This skeleton or core network has much in common with the backbone network for they both maintain the main function of the full network. Since the tinker algorithm is a general method, we look forward to it being put into practical use in future studies.

\section*{APPENDIX A: The details of the tinker algorithm for the ODE model}
The modified tinker algorithm and subroutine $check$ for the ODE model are organized as follows:
\definecolor{shadecolor}{rgb}{0.92,0.92,0.92}
\begin{shaded}
{ \noindent \bf // the C pseudo-code of the tinker algorithm for the ODE model \\
\\ 
tinker($A$, $I$)\{ \\
$~~~$   $I_{remain}$=I, $E_{remain}=\{e_{ij}|a_{ij} \ne 0 , i,j=1,..,N\}$;    \\
$~~~$   do( select node $i$  $\in$  $I_{remain}$  ) \{    \\
$~~~$$~~~$      select $E^0_i=\{e_{ij}|a_{ij} \ne 0, j=1,..,N\}$;    \\
$~~~$$~~~$      mark=flase;  \\
$~~~$$~~~$      for( k=0 ; k $<=$ $d_i$ ; k++) \{    \\
$~~~$$~~~$$~~~$          for( l=1; l $<=$ $C^k_{d_i}$ ; l++) \{    \\
$~~~$$~~~$$~~~$$~~~$         $E_i$=$E_{remain}-E^0_i$+update($E^0_i$, k, l);  //different from previous case    \\
$~~~$$~~~$$~~~$$~~~$         if($check(S^0,E_i)$==ture)mark=true, $E_{remain}=E_i$, break;    \\
$~~~$$~~~$$~~~$          \}    \\
$~~~$$~~~$      if(mark==true)break;  \\
$~~~$$~~~$      \}    \\
$~~~$$~~~$     remove node i from $I_{remain}$;\\
$~~~$   \}while( $I_{remain}$ != $\Phi$ );    \\
$~~~$   $E_{result}$ = $E_{remain}$; $check(S^0,E_{result})$;    \\
\}
}
\end{shaded}

\definecolor{shadecolor}{rgb}{0.92,0.92,0.92}
\begin{shaded}
{\noindent \bf // The subroutine $check$ for the ODE model\\
\bf check($S^0$,$E_i$)\{ \\
$~~~$      flag=true;                          \\
$~~~$      // Use the ODE equations to obtain the time series with \\
$~~~$      // initial condition $S^{0}(0)$ and network $E_i$.  \\
$~~~$      $S(t)$=ode-eq($E_i$,$S^{0}(0)$);             \\
$~~~$      //shift $S(t)$ with proper time $t_0$; \\
$~~~$      $S^{'}(t)=S(t-t_0)$;             \\
$~~~$      // Calculate the Pearson correlation $R$ between $S^{'}$ and $S^{0}$ and the period of $S^{'}$ and $S^{0}$. \\
$~~~$      $[R_i, T_i, T_i^{0},~i=1,...N]$=pc-period($S^{'}, S^{0}$);  \\
$~~~$      for( $i$=1; $i$ $<=$ N; $i$++) \{     \\
$~~~$ $~~~$        if( $R_i<R_0$  OR $|T_i-T_i^{0}|>T_0$ ) flag=false, break;          \\
$~~~$      \}             \\
$~~~$     return flag;    \\
\}
}
\end{shaded}

\section*{APPENDIX B: Boolean network model obtained from the transition of the ODE model}
The Boolean network model of the cell-cycle network of fission yeast is obtained from the transition of the ODE model \cite{Davidich2008b}. The regulatory
network ($A^{'}$) is shown in Fig.~\ref{fig7}(a). Comparing this network with the
original regulatory network ($A$) used in the ODE model, we can get
the following messages: the protein MPF is divided into two parts MPF$1$ and MPF$2$;
M is divided into M and $2$M, where M works at the beginning
of the cell cycle and $2$M plays the role of an indicator for the end
of the cell cycle; Slp$1_{T}$ and Trimer are removed. We use $S_1(t),
S_2(t),..., S_{14}(t)$ to represent the protein states of
Cdc$13_{T}$, preMPF, MPF$1$, MPF$2$, k$_{25}$, k$_{wee}$, M, Slp$1$,
Ste$9$, TF, SK, $2$M, IEP, and Rum$1$, respectively. Here, the Boolean
network model is almost the same with Eq. (\ref{Boolean}) except for a small difference and is described as follows:
\begin{eqnarray}\label{Booleant}
\notag   S_i(t+1)=\left\{\begin{array}{ccccccccc}
      1, &\qquad \text{if}~\sum_{j=1}^{N}a_{ij}^{'}S_j(t)+h_i>0 & \\
      0, &\qquad \text{if}~\sum_{j=1}^{N}a_{ij}^{'}S_j(t)+h_i\leq0, & \\
      \end{array}\right.
\end{eqnarray}
where
\begin{equation}
\notag H=\{h_1, h_2, ..., h_{14}\}=\{0, -1, 0, 0, -0.5, 0.5, 3, 0, 0, 0, 0,
-3, 0, 0.5\}.
\end{equation}
$a_{ij}^{'}=1$ for a green arrow from protein $j$ to $i$ and
$a_{ij}^{'}=-1$ for a pink dashed arrow from protein $j$ to $i$.
The functional sequence $S^{0}(t)$ is shown in
Table \ref{TableS1}.

\begin{table}[!ht] \scriptsize
\begin{center}
\renewcommand\arraystretch{1.0}
\setcounter{table}{0}
\renewcommand{\thetable}{S\arabic{table}}%
\caption{The functional sequences of the cell-cycle process of fission
yeast in the Boolean network model obtained from the transition of the ODE model \cite{Davidich2008b}.}
\begin{tabular}{l|ccccccccccccccc}
\hline
Time &Cdc$13_T$ &preMBF &MBF$1$ &MBF$2$ &k$_{25}$  &k$_{wee}$  &M  &Slp$1$ &Ste$9$ &TF &SK &$2$M &IEP &Rum$1$ \\
\hline t &$S_1^0$ &$S_2^0$ &$S_3^0$ &$S_4^0$ &$S_5^0$ &$S_6^0$ &$S_7^0$
&$S_8^0$ &$S_9^0$ &$S_{10}^0$ &$S_{11}^0$ &$S_{12}^0$ &$S_{13}^0$ &$S_{14}^0$\\
\hline
0  &0   &0   &0   &0   &0   &1   &0   &1   &1   &0   &0   &1   &1   &1\\
1  &0   &0   &0   &0   &0   &1   &1   &1   &1   &1   &0   &0   &0   &1\\
2  &0   &0   &0   &0   &0   &1   &1   &0   &1   &1   &1   &0   &0   &1\\
3  &0   &0   &0   &0   &0   &1   &1   &0   &0   &1   &1   &0   &0   &0\\
4  &1   &0   &0   &0   &0   &1   &1   &0   &0   &1   &1   &0   &0   &0\\
5  &1   &1   &1   &0   &0   &1   &1   &0   &0   &1   &1   &0   &0   &0\\
6  &1   &1   &1   &0   &1   &0   &1   &0   &0   &1   &1   &0   &0   &0\\
7  &1   &0   &1   &0   &1   &0   &1   &0   &0   &1   &1   &0   &0   &0\\
8  &1   &0   &1   &1   &1   &0   &1   &0   &0   &1   &1   &0   &0   &0\\
9  &1   &0   &1   &1   &1   &0   &1   &0   &0   &0   &1   &0   &1   &0\\
10 &1   &0   &1   &1   &1   &0   &1   &1   &0   &0   &0   &0   &1   &0\\
11 &0   &0   &1   &1   &1   &0   &1   &1   &0   &0   &0   &0   &1   &0\\
12 &0   &0   &0   &1   &1   &0   &1   &1   &0   &0   &0   &0   &1   &0\\
13 &0   &0   &0   &0   &0   &1   &1   &1   &1   &0   &0   &0   &1   &1\\
\hline \hline
 \end{tabular}
  \label{TableS1}
  \end{center}
\end{table}

\section*{Acknowledgments}
This study is supported partially by the National Science Foundation of China under
Grant Nos. 11475253, 11405263, and 11675112, the Natural Science Foundation of Zhejiang Province
under Grant No. LY16A050001.


\pagebreak

\pagebreak

\pagebreak

\pagebreak

\pagebreak

\pagebreak

\pagebreak

\pagebreak

\end{document}